# Hereditary properties of permutations are strongly testable[*]

Tereza Klimošová[†]    Daniel Král'[‡]

**Abstract**

We show that for every hereditary permutation property $\mathcal{P}$ and every $\varepsilon_0 > 0$, there exists an integer $M$ such that if a permutation $\pi$ is $\varepsilon_0$-far from $\mathcal{P}$ in the Kendall's tau distance, then a random subpermutation of $\pi$ of order $M$ has the property $\mathcal{P}$ with probability at most $\varepsilon_0$. This settles an open problem whether hereditary permutation properties are strongly testable, i.e., testable with respect to the Kendall's tau distance. In addition, our method also yields a proof of a conjecture of Hoppen, Kohayakawa, Moreira and Sampaio on the relation of the rectangular distance and the Kendall's tau distance of a permutation from a hereditary property.

## 1 Introduction

Property testing is a topic with growing importance with many connections to various areas of mathematics (e.g., see [11, 12, 26] for relation to graph limits) and computer science. A property tester is an algorithm that decides whether a large input object has the considered property by querying only a small sample of it. Since the tester is presented with a part of the input structure, it is necessary to allow an error based on the robustness of the tested property of the input. Following [17, 18], we say that a property $\mathcal{P}$ of combinatorial structures (e.g., graphs) is testable if for every $\varepsilon$, there exists a randomized algorithm $\mathcal{A}$ such that the number of queries made by $\mathcal{A}$ is bounded by a function of $\varepsilon$ independent of the input and such that if the input has the property $\mathcal{P}$, then $\mathcal{A}$ accepts with probability at least $1-\varepsilon$, and if the input is $\varepsilon$-far from $\mathcal{P}$, then $\mathcal{A}$ rejects with probability at least $1-\varepsilon$. The exact notion depends on the studied class $\mathcal{C}$ of combinatorial structures, the considered properties $\mathcal{P}$ and the chosen metric on $\mathcal{C}$. There are also some variants of this basic notion, e.g., one can allow only a one-sided error, i.e., $\mathcal{A}$ is required to accept whenever the input has the property $\mathcal{P}$.

---

[*]The work leading to this invention has received funding from the European Research Council under the European Union's Seventh Framework Programme (FP7/2007-2013)/ERC grant agreement no. 259385.

[†]Computer Science Institute, Faculty of Mathematics and Physics, Charles University, Malostranské náměstí 25, 118 00, Prague, Czech Republic. E-mail: klimosova@iuuk.mff.cuni.cz. This author was also supported by the student grant GAUK 601812.

[‡]Computer Science Institute, Faculty of Mathematics and Physics, Charles University, Malostranské náměstí 25, 118 00, Prague, Czech Republic. E-mail: kral@iuuk.mff.cuni.cz.

The most investigated area of property testing is testing graph properties. One of the most significant results in this area is that of Alon and Shapira [5] asserting that every hereditary graph property, i.e., a property preserved by taking induced subgraphs, is testable, which extends several earlier results [6, 17, 27]. A characterization of testable graph properties can be found in [2]. A logic perspective of graph property testing was addressed in [1, 16] and the connection to graph limits was explored in [26].

Testing properties of other objects have also been intensively studied. For example, results on testing string properties can be found in [3, 7, 24], results related to constraint satisfaction problems in [4] and more algebraically oriented properties are addressed in [8–10, 28–30]. In this paper, we study testing properties of permutations.

To state our results, we need to introduce some terminology. A *permutation of order* $N$ is a bijective mapping from $[N]$ to $[N]$ where $[N]$ denotes the set $\{1,\ldots,N\}$. The order of a permutation $\pi$ is denoted by $|\pi|$. If $\pi$ is a permutation of order $N$ and $X \subseteq [N]$, then the *subpermutation* of $\pi$ *induced* by $X$, denoted by $\pi \upharpoonright X$, is the permutation $\pi'$ of order $|X|$ such that $\pi'(i) < \pi'(j)$ if and only if $\pi(x_i) < \pi(x_j)$ for all $i,j \in [|X|]$ where $X = \{x_1,\ldots,x_{|X|}\}$ and $x_1 < \cdots < x_{|X|}$.

A *permutation property* $\mathcal{P}$ is a set of permutations. If $\pi \in \mathcal{P}$, we say that a permutation $\pi$ has the property $\mathcal{P}$. Since we are interested only in permutation properties in this paper, we often refer to permutation properties just as properties. A property $\mathcal{P}$ is *hereditary* if it is closed under taking subpermutations, i.e., if $\pi \in \mathcal{P}$, then any subpermutation of $\pi$ is in $\mathcal{P}$. An example of a hereditary property is the set of all permutations not containing a fixed permutation as a subpermutation.

There are several notions of distances between permutations, see [15]. The rectangular distance and the Kendall's tau distance will be of most interest to us. Let $\pi$ and $\sigma$ be two permutations of the same order $N$. The *rectangular distance* of $\pi$ and $\sigma$, which is denoted by $\mathrm{dist}_\square(\pi, \sigma)$, is defined as

$$\mathrm{dist}_\square(\pi, \sigma) := \max_{S,T} \frac{||\pi(S) \cap T| - |\pi'(S) \cap T||}{N}$$

where the maximum is taken over all subintervals $S$ and $T$ of $[N]$. The *Kendall's tau distance* $\mathrm{dist}_K(\pi, \sigma)$ is defined as

$$\mathrm{dist}_K(\pi, \sigma) := \frac{|\{(i,j) \text{ such that } \pi(i) < \pi(j), \pi'(i) > \pi'(j) \text{ and } i, j \in [N]\}|}{\binom{N}{2}}.$$

The Kendall's tau distance of two permutations is the minimum number of swaps of consecutive elements transforming $\pi$ to $\sigma$ normalized by $\binom{N}{2}$. Hence, the Kendall's tau distance is considered to correspond to the edit distance of graphs which appears in the hereditary graph property testing, while the rectangular distance is considered to correspond to the cut norm appearing in the theory of graphs limits, see [25]. The latter is demonstrated in the notion of regularity decompositions of permutations developed by Cooper [13, 14] and permutation limits introduced by Hoppen et al. [19, 20] (also see [13, 23] for relation to quasirandom permutations).

If $\mathcal{P}$ is a property, we define the rectangular distance of a permutation $\pi$ from $\mathcal{P}$ as

$$\mathrm{dist}_\square(\pi, \mathcal{P}) = \min_{\sigma \in \mathcal{P}, |\pi|=|\sigma|} \mathrm{dist}_\square(\pi, \sigma)$$

and the Kendall's tau distance to be

$$\mathrm{dist}_K(\pi, \mathcal{P}) = \min_{\sigma \in \mathcal{P}, |\pi|=|\sigma|} \mathrm{dist}_K(\pi, \sigma) \ .$$

It can be shown that if two permutations are close in the Kendall's tau distance, then they are close in the rectangular distance. The converse is not true: the rectangular distance of two random permutation is concentrated around 0 but their Kendall's tau distance is concentrated around $1/2$. Hence, testing permutation properties with respect to the Kendall's tau distance is more difficult than with respect to the rectangular distance (at least in the sense that every tester designed for testing with respect to the Kendall's tau distance also works for testing with respect to the rectangular distance but not vice versa in general). However, one of our results asserts that the converse is true when $\mathcal{P}$ is hereditary.

Testing hereditary permutation properties was first addressed by Hoppen, Kohayakawa, Moreira and Sampaio [21]. They considered testing properties *through subpermutations* where the tester is presented with a random subpermutation of the input permutation (the size of the subpermutation depends on the tested property and the required error). Their main result is the following.

**Theorem 1.** *Let $\mathcal{P}$ be a hereditary property. For any real $\varepsilon > 0$, there exists $M$ such that every permutation $\pi$ of order at least $M$ with $\operatorname{dist}_\square(\pi, \mathcal{P}) > \varepsilon$ satisfies that a random subpermutation of $\pi$ of order $M$ has the property $\mathcal{P}$ with probability at most $\varepsilon$.*

Theorem 1 implies that hereditary properties are testable through subpermutations with respect to the rectangular distance with one-sided error: the tester accepts if the random subpermutation has the property $\mathcal{P}$ and thus the tester always accepts permutations having the property $\mathcal{P}$.

Kohayakawa [22] asked whether hereditary properties of permutations are also testable through subpermutations with respect to the Kendall's tau distance, which he refers to as strong testability. In this paper, we resolve this problem in the positive way. We prove an analogue of Theorem 1 with the rectangular distance replaced with the Kendall's tau distance (Theorem 6). Hence, we establish that hereditary properties are testable through subpermutations with respect to the Kendall's tau distance with one-sided error. Since the Kendall's tau distance for permutations corresponds to the edit distance for graphs, this is viewed in [21] as an analogue of the result of Alon and Shapira [5]. It is also worth noting that, unlike many approaches in this area, our argument is not based on regularity decompositions or on the analysis of limit structures.

Hoppen et al. [21] observed that the strong testability through subpermutation would be implied by the following statement.

**Conjecture 1.** *Let $\mathcal{P}$ be a hereditary property. For every positive real $\varepsilon_0$, there exists $\delta_0$ such that any permutation $\pi$ satisfying $\operatorname{dist}_\square(\pi, \mathcal{P}) < \delta_0$ also satisfies $\operatorname{dist}_K(\pi, \mathcal{P}) < \varepsilon_0$.*

The conjecture is an analogue of the known relation between the rectangular distance and the edit distance to hereditary graph properties from [26].

Our method actually gives the proof of this conjecture which we state as Theorem 7. However, we have decided to include the proof of Theorem 6 instead of just stating that it can be derived from Theorem 7 for completeness.

## 2 Branchings

In this section, we present the notion of branchings which are rooted trees approximately describing hereditary properties. This notion is key in our analysis of hereditary properties.

Let us start with introducing the notion of patterns. If $k$ is an integer, then a *k-pattern* $A$ for an integer $k$ is a sequence $A_1, \ldots, A_\ell$ of non-empty subsets of $[k]$. We refer to $\ell$ as the *length* of $A$ and we write $|A|$ for the length of $A$. The *basic* $k$-pattern is the $k$-pattern of length one comprised of the set $[k]$. A $k$-pattern $A$ is *simple* if each $A_i$ has size one. Finally, a $k$-pattern $A$ is *monotone* if every pair $x \in A_i$ and $x' \in A_{i'}$ with $1 \leq i < i' \leq |A|$ satisfies that $x < x'$.

Before we proceed further, we have to introduce some auxiliary notation. If $A$ is a $k$-pattern, then we write $|A|_i$ for the sum $|A_1| + \cdots + |A_i|$. For completeness, we define $|A|_0 = 0$. If $a$ and $b$ are integers, then $a \bmod b$ is equal to the integer $x \in [b]$ with the same remainder as $a$ after division by $b$.

Fix a $k$-pattern $A$. Let $A_i = \{x_1^i, \ldots, x_{|A_i|}^i\}$ where $x_1^i < \cdots < x_{|A_i|}^i$. For an integer $m$, we define a function $g^{A,m} : [m \cdot |A|_{|A|}] \to [k]$ as

$$g^{A,m}(j) := x^i_{(j - m \cdot |A|_{i-1})) \bmod |A_i|}$$

where $i$ is the largest integer such such that $m \cdot |A|_{i-1} < j$. For example, if $A = \{1, 2, 3\}, \{1, 4\}, \{3\}$, then

$$g^{A,4}(1), \ldots, g^{A,4}(24) = 1, 2, 3, 1, 2, 3, 1, 2, 3, 1, 2, 3, 1, 4, 1, 4, 1, 4, 1, 4, 3, 3, 3, 3\,.$$

Note that the sequence $g^{A,m}(1)g^{A,m}(2) \ldots g^{A,m}(m \cdot |A|_{|A|})$ has $|A|$ blocks such that the $i$-th block consists of $m$ parts each containing the elements of $A_i$ in the increasing order. Besides using the function $g^{A,m}$ here, this function also appears later in the definition of a witnessing pattern and in the proofs of Lemma 3 and Theorem 7.

A permutation $\pi$ is an *m-expansion* of a $k$-pattern $A$ if the following holds:

- the order of $\pi$ is $m \cdot |A|_{|A|}$, and

- if $g^{A,m}(j) < g^{A,m}(j')$ for $j, j' \in \{1, \ldots, m|A|_{|A|}\}$, then $\pi(j) < \pi(j')$.

In other words, if a permutation $\pi$ is an $m$-expansion of $A$, then the range of $\pi$ can be viewed as partitioned into $k$ parts and the permutation $\pi$ consists of $|A|$ groups where the $i$-th group has $m$ blocks of length $|A_i|$ each and the values of $\pi$ in each block belong to the parts of the range of $\pi$ with indices in $A_i$ in the increasing order. The number of $m$-expansions of a $k$-pattern $A$ is equal to

$$\prod_{j=1}^{k} (m \cdot |\{i \text{ such that } i \in [|A|] \text{ and } j \in A_i\}|)! \, .$$

Let $\mathcal{P}$ be a hereditary property. A $k$-pattern $A$ is $\mathcal{P}$-*good* if there exists an $m$-expansion of $A$ in $\mathcal{P}$ for every integer $m$. Otherwise, the pattern $A$ is $\mathcal{P}$-*bad*. So, if $A$ is $\mathcal{P}$-bad, there exists an integer $m$ such that no $m$-expansion of $A$ is in $\mathcal{P}$. The smallest such integer $m$ is called the $\mathcal{P}$-*order* of $A$ and it is denoted by $\langle A \rangle_{\mathcal{P}}$; if $\mathcal{P}$ is clear from the context, we just write $\langle A \rangle$. Observe that if $A$ is $\mathcal{P}$-bad, then no $m$-expansion of $A$ is in $\mathcal{P}$ for every $m \geq \langle A \rangle$.

If $A$ is a $\mathcal{P}$-bad $k$-pattern, then any $k$-pattern $A'$ obtained from $A$ by replacing one element, say $A_i$, by a sequence of at least one and at most $|A_i|\langle A \rangle$ proper subsets of $A_i$ is called a $\mathcal{P}$-*reduction* of $A$. For example, if the 3-pattern $A = \{1\}, \{2, 3\}, \{1, 3\}$ is $\mathcal{P}$-bad and its $\mathcal{P}$-order is two, then one of its $\mathcal{P}$-reductions is $\{1\}, \{2\}, \{2\}, \{3\}, \{1, 3\}$.

The $k$-*branching* of a hereditary property $\mathcal{P}$ is a rooted tree $\mathcal{T}$ such that

- each node $u$ of $\mathcal{T}$ is associated with a $k$-pattern $A^u$,

- the root of $\mathcal{T}$ is associated with the basic $k$-pattern,

- if the pattern $A^u$ of a node $u$ is $\mathcal{P}$-good or simple, then $u$ is a leaf, and

- if the pattern $A^u$ of a node $u$ is $\mathcal{P}$-bad and it is not simple, then the number of children of $u$ is equal to the number of $\mathcal{P}$-reductions of $A$ and the children of $u$ are associated with the $\mathcal{P}$-reductions.

Note that the $k$-branching, i.e., the tree and the association of its nodes with $k$-patterns, is uniquely determined by the property $\mathcal{P}$ and the integer $k$.

Let us argue that the $k$-branching of every hereditary property $\mathcal{P}$ is *finite*. We define the *score* of a $k$-pattern $A$ to be the sequence $m_1, \ldots, m_k$ where $m_i$ is the number of $A_i$'s of cardinality $k + 1 - i$. Observe that the score of a $\mathcal{P}$-reduction of a $\mathcal{P}$-bad $k$-pattern $A$ is always lexicographically smaller than that of $A$. Since the lexicographic ordering on the scores is a well-ordering, the $k$-branching is finite for every hereditary property $\mathcal{P}$.

Let $\mathcal{T}$ be the $k$-branching of a hereditary property $\mathcal{P}$. We now assign to every node $u$ of the $k$-branching of $\mathcal{P}$ an integer weight $w_u$. The weight of a leaf node $u$ is one if $A^u$ is $\mathcal{P}$-good. Otherwise, the weight of a leaf node $u$ is $k\langle A^u\rangle$. If $u$ is an internal node, then $w_u$ is equal to $\langle A^u\rangle km$ where $m$ is the maximum weight of a child of $u$. In particular, the weight of $u$ is at least the weight of any of its children.

## 3 Decompositions

In this section, we introduce a grid-like way of decomposing permutations which we use in our proof. The domain of a permutation will be split into $K$ equal size parts and the range into $k$ such parts with $k \leq K$.

We start with some auxiliary notation. First, $[a/b]_i$ denotes the set of all integers $k \in [a]$ such that $i - 1 < k/\lfloor a/b \rfloor \leq i$. Observe that $|[a/b]_1| = \cdots = |[a/b]_b| = \lfloor a/b \rfloor$ and $|[a/b]_{b+1}| \leq b - 1$. Fix now a permutation $\pi$ of order $N$ and integers $K \in [N]$, $i \in [K]$, $k \in [K]$ and $j \in [k]$. We define $R_{i,j}^{K,k}(\pi)$ as

$$R_{i,j}^{K,k}(\pi) := \{x \in [N/K]_i \text{ such that } \pi(x) \in [N/k]_j\}$$

and we set

$$\rho_{i,j}^{K,k}(\pi) := \frac{|R_{i,j}^{K,k}(\pi)|}{\lfloor N/K \rfloor} \ .$$

Vaguely speaking, $\rho_{i,j}^{K,k}(\pi) \in [0, 1]$ is the density of $\pi$ in the part of the $K \times k$ grid at the coordinates $(i, j)$. If the values of $K$ and $k$ are clear from the context, we will just write $R_{i,j}(\pi)$ and $\rho_{i,j}(\pi)$.

To get used to the definition of the sets $R_{i,j}$ and the quantities $\rho_{i,j}$, we prove an auxiliary lemma which we use later in this section.

**Lemma 2.** *Let $k$ and $K$ be positive integers and let $\varepsilon' \leq 1/(k + 1)$ be a positive real. For every permutation $\pi$ of order at least $k(k + 1)K$ and every $x \in [K]$, there exists $y \in [k]$ such that $\rho_{x,y}(\pi) \geq \varepsilon'$.*

*Proof.* Observe that

$$|R_{x,1}(\pi)| + \cdots + |R_{x,k}(\pi)| \geq \lfloor |\pi|/K \rfloor - k \geq \left(1 - \frac{1}{k+1}\right) \lfloor |\pi|/K \rfloor \ .$$

Since $\varepsilon' \leq 1/(k+1)$, there must exist $y$ such that $\rho_{x,y}(\pi) \geq \varepsilon'$ by the pigeonhole principle. □

Fix a permutation $\pi$, integers $k$, $K$ and $M$ such that $1 \leq k \leq K \leq |\pi|$, and a real $0 \leq \varepsilon' < 1$. If $A$ is a $k$-pattern, then we say that a $K$-pattern $B$ is $(A, M, \varepsilon')$-*approximate* for $\pi$ if the following holds:

- the length of $B$ is $|A|$,

- $B$ is monotone,

- $|B|_{|B|} = \sum_{i=1}^{|B|} |B_i| \geq K - M$, and

- for every $i \in [|A|]$, if $x \in B_i$ and $y \in [k] \setminus A_i$, then $\rho_{x,y}(\pi) < \varepsilon'$.

In other words, an $(A, M, \varepsilon')$-approximate $K$-pattern $B$ decomposes the whole index set $[K]$ except for at most $M$ indices into $|A|$ parts such that the indices contained in the parts determined by $B$ are in the increasing order and for $x \in B_i$, the only dense sets $R_{x,y}(\pi)$ are those with $y \in A_i$.

Suppose that a $k$-pattern $A$ is $\mathcal{P}$-bad for a hereditary property $\mathcal{P}$. We say that a $K$-pattern $B$ is $(A, \varepsilon')$-*witnessing* for $\pi$ if the following holds:

- the length of $B$ is $|A|$,

- there exist integers $1 \leq x_1 < \ldots < x_{|A|_{|A|} \cdot \langle A \rangle} \leq K$ such that $x_j \in B_i$ if $|A|_{i-1} \langle A \rangle < j \leq |A|_i \langle A \rangle$, and

- $\rho_{x_j, g^{A, \langle A \rangle}(j)}(\pi) \geq \varepsilon'$ for every $j \in [|A|_{|A|} \cdot \langle A \rangle]$ (the definition of the function $g$ can be found in Section 2).

In other words, a $K$-pattern $B$ which decomposes the index set $[K]$ is $(A, \varepsilon')$-witnessing, if it is possible to find indices such that there are $|A_i|\langle A \rangle$ indices $x_j$ in each $B_i$ and all the sets $R_{x_j, g^{A, \langle A \rangle}(j)}(\pi)$ are dense. The motivation for this definition is the following: if $B$ is $(A, \varepsilon')$-*witnessing*, then each set $R_{x_j, g^{A, \langle A \rangle}(j)}(\pi)$ has at least $\varepsilon' \lfloor |\pi|/K \rfloor$ elements and consequently at least $(\varepsilon' \lfloor |\pi|/K \rfloor)^{|A|\langle A \rangle}$ subsets of $[|\pi|]$ induce subpermutations that are $\langle A \rangle$-expansions of $A$. This will allow us to deduce that a random subpermutation of sufficiently large order does not have the property $\mathcal{P}$ with high probability.

We now prove a lemma saying that if a $K$-pattern $B$ is approximate but not witnessing with respect to a $k$-pattern $A$ for a permutation $\pi$, then there exists a reduction $A'$ of $A$ and a $K$-pattern $B'$ such that $B'$ is approximate with respect to $A'$.

**Lemma 3.** *Let $\mathcal{P}$ be a hereditary property, let $k$, $K$, $m$ and $M$ be positive integers and let $\varepsilon' \leq 1/(k+1)$ be a positive real. Suppose that a $\mathcal{P}$-bad $k$-pattern $A$ and a monotone $K$-pattern $B$ with $|A| = |B|$. If the pattern $B$ is $(A, M, \varepsilon')$-approximate for a permutation $\pi$, $|\pi| \geq k(k+1)K$, $B$ is not $(A, \varepsilon')$-witnessing for $\pi$ and $|B_i| \geq mk\langle A\rangle$ for every $i \in [|B|]$, then there exist a $\mathcal{P}$-reduction $A'$ of $A$ and a monotone $K$-pattern $B'$ such that*

- *the lengths of $A'$ and $B'$ are the same,*

- *$B'$ is $(A', M + mk\langle A\rangle, \varepsilon')$-approximate for $\pi$, and*

- *$|B'_i| \geq m$ for every $i \in [|B'|]$.*

*Proof.* If $B$ is not $(A, \varepsilon')$-witnessing for $\pi$, then there exists an index $j \in [|B|]$ such that there is no $|A_j|\langle A\rangle$-tuple $x_1 < \cdots < x_{|A_j|\langle A\rangle}$ in $B_j$ satisfying $\rho_{x_i,y_i}(\pi) \geq \varepsilon'$ where $y_i = g^{A,\langle A\rangle}(|A|_{j-1}\langle A\rangle + i)$. Fix such an index $j$ for the rest of the proof.

If $|A_j| = 1$, then an $\langle A\rangle$-tuple with the properties given in the previous paragraph is formed by any $\langle A\rangle$ elements of $B_j$ by Lemma 2. So, we assume that $|A_j| \geq 2$ in the rest of the proof. Define $x_1$ to be the smallest index in $B_j$ such that $\rho_{x_1,y_1}(\pi) \geq \varepsilon'$. Suppose that we have defined the indices $x_1, \ldots, x_i$ and define $x_{i+1}$ to be the smallest index in $B_j$ that is larger than $x_i$ such that $\rho_{x_{i+1},y_{i+1}}(\pi) \geq \varepsilon'$. If no such index exists, we stop constructing the sequence. Let $\ell$ be the number of the indices defined. By the choice of $j$, $\ell < |A_j|\langle A\rangle$. For completeness, set $x_0 = 0$ and $x_{\ell+1} = K + 1$.

Define $C_i$, $i \in [\ell+1]$, to be the set of the elements of $B_j$ strictly between $x_{i-1}$ and $x_i$. If the subset $C_i$ has size less than $m$, remove it from the sequence and let $C'_1, \ldots, C'_{\ell'}$ be the resulting sequence. Observe that

$$\begin{aligned}|B_j| - \sum_{i=1}^{\ell'}|C'_i| &\leq \ell + (\ell+1)(m-1) \leq (\ell+1)m - 1\\ &\leq m|A_j|\langle A\rangle - 1 \leq mk\langle A\rangle - 1\end{aligned} \quad (1)$$

since the sets $C'_1, \ldots, C'_{\ell'}$ contain all the elements of $B_j$ except for the elements $x_1, \ldots, x_\ell$ and the elements contained in the sets $C_1, \ldots, C_{\ell+1}$ with cardinalities at most $m - 1$. In particular, we can infer from $|B_j| \geq mk\langle A\rangle$ that $\ell' \geq 1$.

Next, define $C''_i$, $i \in [\ell']$, to be the set of $y \in [k]$ such that there exists $x \in C'_i$ with $\rho_{x,y}(\pi) \geq \varepsilon'$. Lemma 2 implies that the sets $C''_1, \ldots, C''_{\ell'}$ are non-empty. We infer from the way we have chosen the indices $x_1, \ldots, x_\ell$ that each set $C_i''$ is a proper subset of $A_j$. Finally, define the $k$-pattern $A'$ to be the $K$-pattern $A$ with $A_j$ replaced with $C_1'', \ldots, C_{\ell'}''$ and the $K$-pattern $B'$ to be the $K$-pattern $B$ with $B_j$ replaced with $C_1', \ldots, C_{\ell'}'$. By the definition of $C_1'', \ldots, C_{\ell'}''$ and by (1), the $K$-pattern $B'$ is $(A', M + mk\langle A\rangle, \varepsilon')$-approximate for $\pi$. By the choice of $C_1', \ldots, C_{\ell'}'$, we have that $|B_i'| \geq m$ for every $i \in [|B'|]$. Finally, since $\ell' \leq \ell \leq |A_j|\langle A\rangle$ and every $C_i'$, $i \in [\ell']$, is a proper subset of $B_j$, $A'$ is $\mathcal{P}$-reduction of $A$. □

We finish this section with the following lemma on approximating the structure of a sufficiently large permutation $\pi$ with respect to a hereditary property.

**Lemma 4.** *Suppose $\mathcal{P}$ is a hereditary property. For all integers $k$ and reals $\varepsilon$ and $\varepsilon'$, $0 < \varepsilon \leq 1$ and $0 < \varepsilon' \leq 1/(k+1)$, there exists an integer $K$ such that for every permutation $\pi$ of order at least $k(k+1)K$, there exist a $k$-pattern $A$ and a $K$-pattern $B$ with the same lengths such that*

- *$A$ is $\mathcal{P}$-bad and $B$ is $(A, \varepsilon')$-witnessing for $\pi$, or*

- *$A$ is $\mathcal{P}$-good and $B$ is $(A, \lfloor\varepsilon K\rfloor, \varepsilon')$-approximate for $\pi$.*

*Proof.* Let $\mathcal{T}$ be the $k$-branching with respect to $\mathcal{P}$. Let $d$ be the depth of $\mathcal{T}$, i.e., the maximum number of vertices on a path from the root to a leaf, and let $w_0$ be the weight of the root of $\mathcal{T}$. We show that $K := \lceil dw_0/\varepsilon\rceil$ has the properties claimed in the statement of the lemma.

Let $\pi$ be a permutation of order at least $k(k+1)K$. Based on $\pi$, we define a path from the root to one of the nodes in $\mathcal{T}$ in a recursive way. In addition to choosing the nodes $u^i$ on the path, we also define monotone $K$-patterns $B^i$ such that $B^i$ is $(A^{u^i}, i \cdot w_0, \varepsilon')$-approximate for $\pi$ and $|B_j^i| \geq w_{u^i}$ for every $j \in [|B^i|]$.

Let $u_0$ be the root of $\mathcal{T}$ and set $B^0$ to be the basic $K$-pattern. Clearly, $B^0$ is $(A^{u^0}, 0, \varepsilon')$-approximate for $\pi$. Suppose that the node $u^i$ on the path has already been chosen and we now want to choose the next node. If $u^i$ is a leaf node, we stop. If $u^i$ is not a leaf node, then the $k$-pattern $A^{u^i}$ must be $\mathcal{P}$-bad. If $B^i$ is $(A, \varepsilon')$-witnessing for $\pi$, we also stop. Otherwise, Lemma 3 applied with $m$ equal to the maximum weight of a child of $u^i$ (note that $|B_j^i| \geq mk\langle A^{u^i}\rangle$ for every $j \in [|B^i|]$) implies that there exist a $\mathcal{P}$-reduction $A'$ of $A^{u^i}$ and a $K$-pattern $B^{i+1}$ such that $B^{i+1}$ is $(A', i \cdot w_0 + mk\langle A^{u^i}\rangle, \varepsilon')$-approximate for $\pi$ and $|B_j^{i+1}| \geq m$ for every $j \in [|B^{i+1}|]$. Choose $u^{i+1}$ to be the child of $u^i$ such that $A^{u^{i+1}} = A'$. Since $mk\langle A^{u^i}\rangle \leq w_0$, we obtain that $B^{i+1}$ is $(A^{u^{i+1}}, (i+1)w_0, \varepsilon')$-approximate for $\pi$.

Let $\ell$ be the length of the constructed path. We claim that the $k$-pattern $A^{u^\ell}$ and the $K$-pattern $B^\ell$ have the properties described in the statement of the lemma.

If $u^\ell$ is not a leaf node, then $A^{u^\ell}$ is $\mathcal{P}$-bad and $B^\ell$ is $(A^{u^\ell}, \varepsilon')$-witnessing for $\pi$ (since we have stopped at $u^\ell$). If $u^\ell$ is a leaf node and $A^{u^\ell}$ is $\mathcal{P}$-bad, then $B^\ell$ is $(A^{u^\ell}, \varepsilon')$-witnessing for $\pi$ by Lemma 3 applied for $m = 1$ ($A^{u^\ell}$ cannot have a $\mathcal{P}$-reduction because it is simple). Finally, if $u^\ell$ is a leaf node and $A^{u^\ell}$ is $\mathcal{P}$-good, $B^\ell$ is $(A^{u^\ell}, \lfloor \varepsilon K \rfloor, \varepsilon')$-approximate for $\pi$ since $dw_0 \leq \lfloor \varepsilon K \rfloor$. □

## 4 Testing

In this section, we establish our main result. The next lemma, which says that every permutation that is far from a hereditary property $\mathcal{P}$ in the Kendall's tau distance has a witnessing $K$-pattern for a suitable choice of $k$ and $K$, is the core of our proof.

**Lemma 5.** *Let $\mathcal{P}$ be a hereditary property. For every real $\varepsilon_0 > 0$, there exist integers $k$, $K$ and $M$, and a real $\varepsilon' > 0$ such if $\pi$ is a permutation of order at least $M$ with $\mathrm{dist}_K(\pi, \mathcal{P}) \geq \varepsilon_0$, then there exist a $\mathcal{P}$-bad $k$-pattern $A$ and a $K$-pattern $B$ with the same length such that $B$ is $(A, \varepsilon')$-witnessing for $\pi$.*

*Proof.* Without loss of generality, we can assume that $\varepsilon_0 < 1$. Set $k = \lceil 10/\varepsilon_0 \rceil$, $\varepsilon = \varepsilon_0/10$ and $\varepsilon' = \varepsilon_0/(10k+10) \leq 1/(k+1)$. Let $K$ be the integer from the statement of Lemma 4 applied for $\mathcal{P}$, $k$, $\varepsilon$ and $\varepsilon'$. Using this value, set

$$M = \max\left\{ k(k+1)K, \left\lceil \frac{10k}{\varepsilon_0} \right\rceil, \left\lceil \frac{10K}{\varepsilon_0} \right\rceil \right\}.$$

We show that this choice of $k$, $K$, $M$ and $\varepsilon'$ satisfies the assertion of the lemma.

Let $\pi$ be a permutation of order $N \geq M$. Apply Lemma 4 to $\pi$. Let $A$ be the $k$-pattern and $B$ the $K$-pattern as in the statement of the lemma. Either $A$ is $\mathcal{P}$-bad and $B$ is $(A, \varepsilon')$-witnessing for $\pi$, which is the conclusion of the lemma, or $A$ is $\mathcal{P}$-good and $B$ is $(A, \varepsilon K, \varepsilon')$-approximate for $\pi$. Hence, we assume the latter and deduce that $\mathrm{dist}_K(\pi, \mathcal{P}) < \varepsilon_0$.

To reach our goal, we define two auxiliary functions $f_B : [N] \to [|B|]$ and $f_A : [N] \to [k]$. Informally speaking, when searching for a permutation in $\mathcal{P}$ close to $\pi$, we consider an $m$-expansion of $A$ for a very large integer $m$ and we show that one of its subpermutations is close to $\pi$. As explained after the definition of an $m$-expansion, every $m$-expansion can be viewed as consisting of $|A| = |B|$ blocks where the $i$-th block has $m \cdot |A_i|$ elements. In the subpermutation we construct, we choose the element corresponding to $x \in [N]$ in the $f_B(x)$-th block of an $m$-expansion of $A$ and the value of $g^{A,m}$ for this element will be the $f_A(x)$-th smallest element of $A_{f_B(x)}$.

Let us now proceed in a formal way. First, we define the function $f_B$. Let $x \in [N]$ and let $i$ be the integer such that $x \in [N/K]_i$. Let $j$ be the largest integer such that $i$ is smaller than all the elements of $B_j$; if no such set exists, let $j = |B| + 1$. Set $f_B(x) := \max\{1, j-1\}$. Clearly, $f_B$ is non-decreasing and if $i \in B_j$, then $f_B(x) = j$ for every $x \in [N/K]_i$. We now proceed with defining the function $f_A$. If $i \in B_{f_B(x)}$, $\pi(x) \in [N/k]_{i'}$ such that $i' \in [k]$ and $\rho_{i,i'}(\pi) \geq \varepsilon'$, set $f_A(x) = i''$ where $i''$ is the number of elements of $A_{f_B(x)}$ smaller or equal to $i'$. Otherwise, set $f_A(x) = 1$.

Since $A$ is $\mathcal{P}$-good, there exists an $N$-expansion $\sigma$ of $A$ that is in $\mathcal{P}$. Set

$$z_x := |A|_{f_B(x)-1} N + x|A_{f_B(x)}| + f_A(x) \quad \text{for } x \in [N].$$

Observe that $1 \leq z_1 < \cdots < z_N \leq N \cdot |A|_{|A|}$. In the rest of the proof, we establish that the subpermutation $\pi'$ of $\sigma$ induced by $\{z_1, \ldots, z_N\}$ satisfies $\mathrm{dist}_K(\pi, \pi') \leq \varepsilon_0$. Since $\mathcal{P}$ is hereditary and $\sigma \in \mathcal{P}$, this implies $\mathrm{dist}_K(\pi, \mathcal{P}) \leq \varepsilon_0$.

We now define five types of pairs $(x, x')$, $1 \leq x < x' \leq N$. Suppose that $x \in [N/K]_i$, $\pi(x) \in [N/k]_j$, $x' \in [N/K]_{i'}$ and $\pi(x') \in [N/k]_{j'}$.

- The pair $(x, x')$ is of *Type I* if $i = K+1$ or $i' = K+1$.

- The pair $(x, x')$ is of *Type II* if $j = k+1$ or $j' = k+1$.

- The pair $(x, x')$ is of *Type III* if it is not of Type I and $i \notin B_{f_B(x)}$ or $i' \notin B_{f_B(x')}$.

- The pair $(x, x')$ is of *Type IV* if it is neither of Type I nor of Type II, and $\rho_{i,j} < \varepsilon'$ or $\rho_{i',j'} < \varepsilon'$.

- The pair $(x, x')$ is of *Type V* if it is not of Type II and $j = j'$.

We now estimate the number of pairs $(x, x')$, $1 \leq x < x' \leq N$, of each of the five types. The number of pairs of Type I is at most $K(N-1) \leq$ $\varepsilon_0 N(N-1)/10$ since $|[N/K]_{K+1} \cap [N]| \leq K$. Similarly, the number of pairs of Type II is at most $k(N-1) \leq \varepsilon_0 N(N-1)/10$ since $|[N/k]_{k+1} \cap [N]| \leq k$. The number of pairs of Type III is at most $\varepsilon N(N-1) = \varepsilon_0 N(N-1)/10$ since $K - (|B_1| + \cdots + |B_{|B|}|) \leq \varepsilon K$.

For $i \in [K]$ and $j \in [k]$ with $\rho_{i,j}(\pi) < \varepsilon'$, the number of the choices of $x \in [N/K]_i$ with $\pi(x) \in [N/k]_j$ is at most $\varepsilon' N/K$. Hence, the number of $x$ with this property for some $i$ and $j$ is at most $\varepsilon' kN < \varepsilon_0 N/10$. Consequently, the number of pairs of Type IV is strictly less than $\varepsilon_0 N(N-1)/10$. Finally, for $x$ with $\pi(x) \in [N/k]_j$, the number of choices of $x' \neq x$ with $\pi(x') \in [N/k]_j$ is at most $N/k - 1$. Hence, the number of pairs of Type V is strictly less than $N(N/k - 1) \leq N(N-1)/k \leq \varepsilon_0 N(N-1)/10$.

We conclude that the number of pairs $(x, x')$, $1 \leq x < x' \leq N$, that are of at least of one of Types I–V is at most $\varepsilon_0 N(N-1)/2$.

We claim that if the pair $(x, x')$, $1 \leq x < x' \leq N$, is not of any of the Types I–V, then $\pi(x) < \pi(x')$ if and only if $\pi'(x) < \pi'(x')$. Let $i$, $i'$, $j$ and $j'$ be chosen as in the previous paragraph. Suppose $\pi(x) < \pi(x')$. If $(x, x')$ is not of any of the Types I–V, then it holds that $i \in B_{f_B(x)}$, $i' \in B_{f_B(x')}$, $j \neq j'$, $\rho_{i,j}(\pi) \geq \varepsilon'$ and $\rho_{i',j'}(\pi) \geq \varepsilon'$. This implies that the $f_A(x)$-th smallest element of $A_{f_B(x)}$ is smaller than the $f_A(x')$-th smallest element of $A_{f_B(x')}$. Consequently, $\pi'(x) < \pi'(x')$ by the choice of $z_x$ and $z_{x'}$. Analogously, one can show that if $\pi(x) > \pi(x')$, then $\pi'(x) > \pi'(x')$.

Since the number of pairs $(x, x')$, $1 \leq x < x' \leq N$, of at least one of the five types is at most $\varepsilon_0 N(N-1)/2$, we get that $\text{dist}_K(\pi, \pi') < \varepsilon_0$ as desired. □

We are now ready to prove our main theorem. Note that Theorem 6 implies that hereditary properties of permutations are strongly testable through subpermutations: for $\varepsilon > 0$, the tester take a random subpermutation of order $M_0$ from the statement of Theorem 6 and it accepts if the random subpermutation has the tested property.

**Theorem 6.** *Let $\mathcal{P}$ be a hereditary property. For every positive real $\varepsilon_0$, there exists $M_0$ such that if $\pi$ is a permutation of order at least $M_0$ with $\text{dist}_K(\pi, \mathcal{P}) \geq \varepsilon_0$, then a random subpermutation $\pi$ of order $M_0$ has the property $\mathcal{P}$ with probability at most $\varepsilon_0$.*

*Proof.* Without loss of generality, we assume that $\varepsilon_0 < 1$. Apply Lemma 5 to $\mathcal{P}$ and $\varepsilon_0$. Let $k$, $K$ and $M$ be the integers and let $\varepsilon'$ be the real as in the statement of the lemma. Note that we can also assume that $\varepsilon' < 1$. Set $M_0$ as
$$M_0 = \max\left\{M, K(K+1), \frac{\log\frac{kK}{\varepsilon_0}}{\log\frac{K+1}{K+1-\varepsilon'}}\right\}.$$

Let $\pi$ be a permutation of order $N \geq M_0$. Note that the probability that a random $M_0$-element subset $X$ of $[N]$ contains no element of a set $R_{i,j}(\pi)$ with $\rho_{i,j}(\pi) \geq \varepsilon'$ is at most
$$\left(1 - \frac{|R_{i,j}(\pi)|}{N}\right)^{M_0} = \left(1 - \rho_{i,j}(\pi)\left\lfloor\frac{N}{K}\right\rfloor\frac{1}{N}\right)^{M_0} \leq \left(1 - \frac{\varepsilon'}{K+1}\right)^{M_0} \leq \frac{\varepsilon_0}{kK}.$$

By the union bound, the probability that there exists $i \in [K]$ and $j \in [k]$ with $\rho_{i,j}(\pi) \geq \varepsilon'$ such that $X$ contains no element from the set $R_{i,j}(\pi)$ is at most $\varepsilon_0$. This implies that with probability at least $1 - \varepsilon_0$ a random $M_0$-element subset $X$ of $[M_0]$ contains at least one element from each set $R_{i,j}(\pi)$ with $\rho_{i,j}(\pi) \geq \varepsilon'$.

By Lemma 5, if $\text{dist}_K(\pi, \mathcal{P}) \geq \varepsilon_0$, there exists a $k$-pattern $A$ and a $K$-pattern $B$ such that $A$ is $\mathcal{P}$-bad and $B$ is $(A, \varepsilon')$-witnessing for $\pi$. Since a random $M_0$-element subset of $[N]$ contains an element from each $R_{i,j}(\pi)$ with $\rho_{i,j}(\pi) \geq \varepsilon'$ with probability at least $1 - \varepsilon_0$, a random $M_0$-element subpermutation of $\pi$ contains an $\langle A \rangle$-expansion of $A$ as a subpermutation with probability at least $1 - \varepsilon_0$. Consequently, a random $M_0$-element subpermutation of $\pi$ is not in $\mathcal{P}$ with probability at least $1 - \varepsilon_0$. □

We are also in a position to prove that, for hereditary properties $\mathcal{P}$, the function $\text{dist}_K(\pi, \mathcal{P})$ is continuous with respect to the metric given by $\text{dist}_\square$ in the sense considered in [21].

**Theorem 7.** *Let $\mathcal{P}$ be a hereditary property. For every $\varepsilon_0 > 0$, there exists $\delta_0 > 0$ such that any permutation $\pi$ satisfying $\text{dist}_\square(\pi, \mathcal{P}) < \delta_0$ also satisfies $\text{dist}_K(\pi, \mathcal{P}) < \varepsilon_0$.*

*Proof.* Apply Lemma 5 to $\mathcal{P}$ and $\varepsilon_0$. Let $k$, $K$ and $M$ be the integers and let $\varepsilon'$ be the real as in the statement of the lemma. Set $M_0$ to be the maximum of $M$ and $K+1$, and set $\delta_0$ to be the minimum of $1/M_0$ and $\frac{\varepsilon'}{4K}$.

Suppose that there exists $\sigma \in \mathcal{P}$ with $|\pi| = |\sigma|$ and $\text{dist}_\square(\pi, \sigma) < \delta_0$. If the order of $\pi$ is smaller than $M_0$, then $\pi$ and $\sigma$ must be the same which yields $\text{dist}_\square(\pi, \mathcal{P}) = \text{dist}_K(\pi, \mathcal{P}) = 0$. So, we can assume that the order of $\pi$ is at least $M_0$.

Assume to contrary that $\operatorname{dist}_K(\pi, \mathcal{P}) \geq \varepsilon_0$. By Lemma 5, there exists a $\mathcal{P}$-bad $k$-pattern $A$ and a $K$-pattern $B$ such that $B$ is $(A, \varepsilon')$-witnessing for $\pi$. By the choice of $\delta_0$, $B$ is $(A, \varepsilon'/2)$-witnessing for $\sigma$ (recall that the order of $\pi$ is at least $K + 1$). This yields that $R_{x_j, g^{A, \langle A \rangle}(j)}(\sigma) \neq \emptyset$ for every $j \in [|A|_\ell \cdot \langle A \rangle]$ where $x_j$ are chosen as in the definition of $(A, \varepsilon'/2)$-witnessing. In particular, $\sigma$ contains a subpermutation not in $\mathcal{P}$ (choose one element from each of the sets $R_{x_j, g^{A, \langle A \rangle}(j)}$ and consider the subpermutation induced by the chosen elements) which is impossible since $\mathcal{P}$ is hereditary. □